# Separation of Single walled Carbon Nanotubes by Light

*Madhusudana Gopannagari[†], Vikram Bakaraju[†], Harsh Chaturvedi[†*]*

[†]Department of Physics, Indian Institute of Science Education and Research, Pune-411008, INDIA

ABSTRACT: We report optical separation of metallic and semiconducting single walled carbon nanotubes of specific diameters from well-dispersed, stable solution. This phenomenon of optical separation was studied using lamps of different frequencies across UV, Vis and NIR spectrum. Optically aggregated SWNTs were separated from the SWNTs remaining dispersed in the solution and characterized using absorption and Raman spectroscopy. Separated SWNTs distinctly show enrichment of metallic SWNTs using UV lamp, whereas different diameter of SWNTs aggregates out depending on the frequency of illumination. Fabricated field effect transistors based on the separated SWNTs under UV lamp confirms enrichment of metallic SWNTs in the supernatant where as aggregated SWNTs shows semiconducting behavior. Spectral changes and electronic (IV) measurements of our optically separated SWNTs were compared and verified with the reference solution of pre-separated, pure metallic and semiconducting SWNTs.

[*] Corresponding Author E-mail: hchaturv@iiserpune.ac.in

Single walled carbon nanotubes (SWNTs) are being widely researched and proposed as important nanomaterial for novel electro-optic devices, sensors.[1, 2] Electronic property of nanotubes depends on the diameter of the tube and SWNTs as batch produced are mixture of nanotubes of varying diameters, thus having both semiconducting and metallic properties.[3-5] For any potential application based on pristine or functionalized SWNTs, it is of immense importance to be sure of the electronic type of the tubes. Sensors[2], logic gates[6] and other electro-optical applications[1] based on field effect transistor need semiconducting SWNTs; whereas for several other applications, such as for optics and plasmonics metallic SWNTs are preferred.[7] Therefore, it is important to develop technologies to commercially separate semiconducting and metallic nanotubes. Various separation protocols including electromagnetic or chemical methods have been reported.[8-13] However, these methods have demonstrated their own limitations such as scalability to microgram levels, creation of defects, adding of functional groups resulting in undesirable changes to the pristine nature of SWNTs. Except for few groups which are using surfactant and chemical modifications to efficiently separate metallic and semiconducting SWNTs; not much advances have been made for large scale, efficient separation of different diameter of these nanotubes from pristine SWNTs. Smith, et al.[14] theoretically predict photophoretic separation in chiral single-walled carbon nanotubes by resonant optical sorting technique. Certainly, there is an urgent need to develop efficient optical processes devoid of chemical modifications and can overcome above said limitations.

Dispersed solution of SWNTs provides us with experimental opportunity to study the effects of electro-optical forces on the intermolecular interactions; thus affecting colloidal stability of the nanoparticles. Along with the photophoresis of pristine SWNT aggregates, the phenomenon of optically induced aggregation of pristine SWNTs from well dispersed stable solution has recently been reported by us. Where in we have shown that optical illumination significantly affects the colloidal stability, leading to aggregation in specific SWNTs. Rate of this optically induced aggregation of pure SWNTs, has been shown to depend on the frequency and intensity of illumination, type of solvent and also on the type of SWNTs (semiconducting/metallic). Using the reported phenomenon as a technique, here in we demonstrate potential large scale process for separation of different diameter of SWNTs by light. Dispersed SWNTs were separated from optically aggregated SWNTs under different lamps and characterized using absorption Raman spectroscopy and electro-optical measurements. Devices were fabricated using our optically separated SWNTs to assess the commercial and practical viability of the process for electro-optical applications. Moreover, we have also characterized and compared our optically separated SWNTs with the reference solution of metallic and semiconducting SWNTs.

Well dispersed, stable solutions of pristine SWNTs were prepared by well reported method of ultrasonication of 0.6 mg of SWNT (Nano Integris®-IsoNanotubes, 95% purity) in 100 mL of N,N-Dimethylformamide (Sigma Aldrich®).[15] SWNT solutions were found to be kinetically stable for several weeks. Prepared, stable solutions of well-dispersed, pristine SWNTs in a Spectrosil quartz cuvette were exposed to UV lamp (125 W, 352 nm), Mercury lamp of broadband UV-visible frequency (80 Watts), Sodium light source (Na) lamp and NIR lamp (Fuzi, 150Watts) for 90 minutes (**Figure 1a**). Samples were then centrifuged @ 8000 rpm (Eppendorf® Mini Spin Centrifuge Machine) for 10 min to carefully separate optically aggregated SWNTs, from the SWNTs remaining dispersed in the solution. Separated supernatant and aggregated SWNTs were analyzed by UV-Vis-NIR spectrophotometer (Perkin Elmer® Lambda 950 UV-Vis spectrometer) and Raman Spectroscopy using He-Ne laser operating at 632 nm (LabRAM). **Figure 1(a)** shows, images of quartz cuvette showing optically induced aggregation of pristine SWNTs (Right) from the well dispersed solution (Left), after exposure to different lamps for 90 minutes. Aggregated floc as observed in each of the cuvettes, exposed to different lamps are indicated in the picture **Figure 1(a)**.

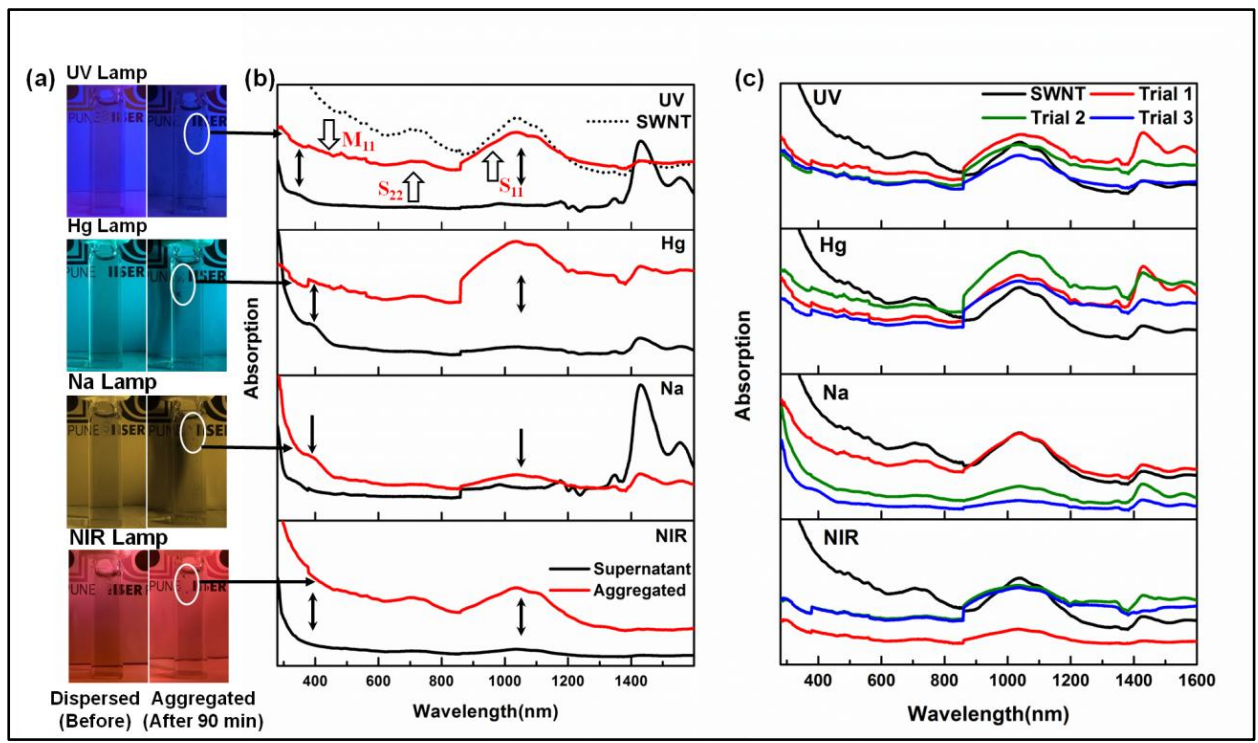

**Figure 1.** (a) Images showing the dispersed and aggregated samples of SWNT after 90 minutes of illumination under UV, Hg, Na and NIR lamps. (b) Absorption spectra for supernatant and aggregates separated in pristine SWNTs using frequency of light. UV-Vis absorption band indicates specific separation of SWNTs due to optical excitation of different frequencies. (c) Absorption spectra of aggregated pristine SWNTs collected after 30, 60 and 90 minutes light illumination.

**Figure 1(b)** shows absorption spectra of the pristine SWNTs in UV-Vis-NIR region (250 nm-1600 nm). Discrete peaks (van Hove singularities) in visible NIR band and background plasmonic absorption in the UV frequency are observed due to both semiconducting and metallic

SWNTs, mixed in the pristine SWNT solution. Absorption peaks in the NIR frequency, as indicated in the **Figure 1(b)**, corresponds to the first optical transition $S_{11}$ (900-1600 nm), band gap absorption of the semiconducting SWNTs.[16] Similarly the optical transitions between the second energy $S_{22}$ (550-900 nm) bands are observed in the visible region. However, enhanced, exponential, background absorption in the UV frequency (350-600 nm), as observed in the **Figure 1(b)** is attributed to the plasmonic resonance by metallic SWNTs.

**Figure 1(b)** shows the absorption spectra for optically separated SWNTs under different lamps (UV, Hg, Na, NIR). Separated SWNTs, both the ones optically aggregated and SWNTs remaining dispersed in the supernatant solution show complimentary, discrete changes in the absorption spectra. Changes in the plasmonic band due to metallic SWNTs has been indicated by the arrows in **Figure 1(b)**. Moreover, changes in the NIR band which corresponds to the band gap absorption by semiconducting SWNTs also show significant changes, as indicated in the **Figure 2(a)**. As compared to the absorption spectra of pristine SWNTs, SWNTs which aggregated under UV lamp shows enhanced absorption in the Vis -NIR region (500-1200nm) where as SWNTs remaining dispersed in supernatant solution, show enhanced plasmonic absorption (300-450 nm) due to metallic SWNTs.

Similarly, under mercury lamp SWNTs separated in the supernatant show wider plasmonic absorption in UV and visible frequencies. Correspondingly aggregated SWNTs show discrete increase in certain van Hove singularities (400-700 nm) indicating enrichment in specific diameter of semiconducting SWNTs. SWNTs in the separated supernatant, under Na lamp show distinct absorption around 400 nm, complimentarily SWNTs in the aggregated floc shows discrete absorption in the vis-NIR region . Absorption spectra of the SWNTs separated under NIR lamp also show, unambiguous changes between the supernatant and the aggregated floc, as compared to pristine SWNTs; indicating enhanced absorption at certain frequencies due to enrichment in those specific diameter of SWNTs. **Figure 1(c)** shows absorption spectra of SWNTs in the supernatant, separated after 30, 60 and 90 minutes of optical illumination by different lamps. Consistent and similar changes are observed in the absorption spectra of the separated SWNTs exposed to varying time intervals, indicating aggregation of selective SWNTs only.

**Figure 2** compares, NIR absorption and RBM Raman spectra of pristine-SWNTs separated under UV, mercury, sodium and NIR illumination. Since absorption by SWNTs in the NIR frequency is directly related to the band gap absorption by semiconducting tubes, significant increase in the NIR absorption in separated aggregates indicates enrichment of semiconducting SWNTs. Diameter ($d_t$) of the semiconducting SWNTs is calculated using the absorption in the NIR frequency ($S_{11}$) by $d_t = \frac{S_{11}(2a_{c-c}\gamma_0)}{hc}$. Where, $a_{c-c}$ is c-c bond length (0.142 nm), and $\gamma_0$=2.7 eV is the c-c tight binding constant, h is the plank's constant and c is the velocity of light in vaccum.[16] Variations in the NIR peak intensity and frequency can be associated with the changes in the relative distribution of the semiconducting SWNTs as shown in **Figure 2(a)**.

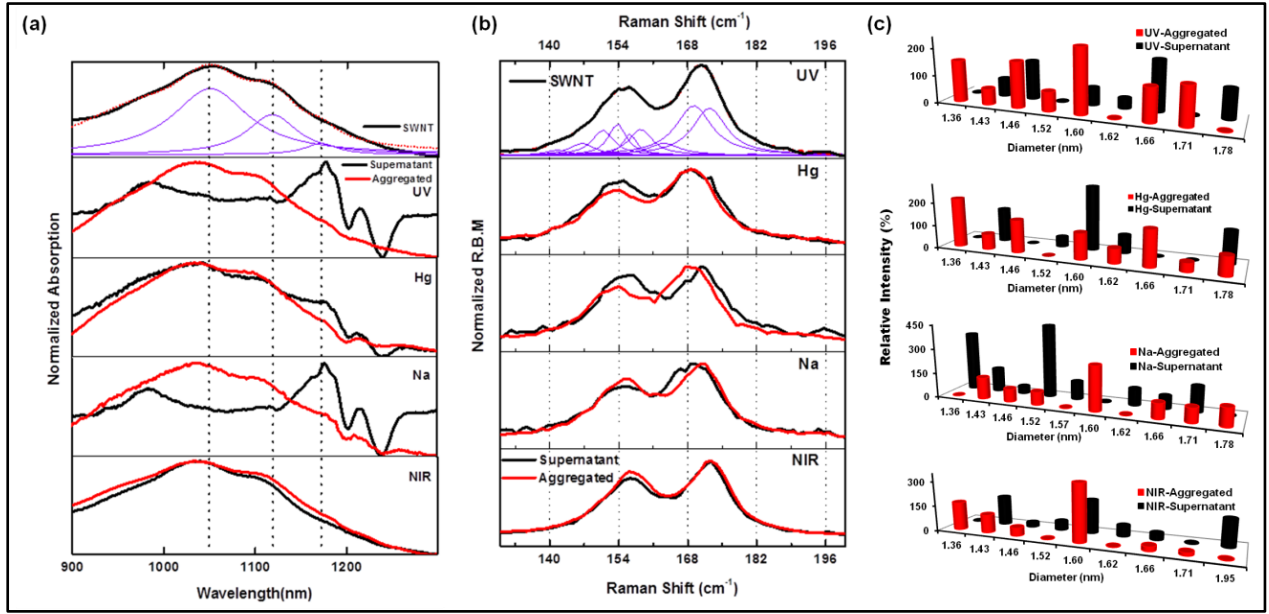

**Figure 2.** (a) NIR bands show specific SWNTs are separating from supernatant and aggregated (b) Raman spectra of RBM modes of SWNTs illuminated with UV, Hg, Na and NIR lamps. (c) Relative histogram of diameter calculation from radial breathing modes.

Diameter of individual SWNTs is related to the frequency of RBM peak by $\omega_{RBM} = (\alpha_{RBM}/d) + \alpha_{bundle}$.[17] Where, $\alpha_{RBM}, \alpha_{bundle}$ are constants and $d$ is the diameter of the SWNT corresponding to the RBM peak frequency($\omega_{RBM}$). Radial breathing mode (RBM), Raman spectra of the separated nanotubes shown in **Figure 2(b)**, also shows discernible changes indicating enrichment in specific diameter of SWNTs due to optical aggregation under UV, Hg, Na and NIR lamps. RBM of each sample was fit using multiple lorentzian functions to identify relative enrichment in individual diameter tubes in the Raman spectra of pristine SWNT. Intensity of RBM peaks in each separated SWNTs sample was normalized with respect to the RBM peaks of un-separated, pristine SWNT sample. Individual RBM of each sample was fitted and normalized using multiple lorentzian peaks, representing different diameter of SWNTs. As calculated from the expression, RBM of aggregates and supernatant shows relative enrichment under the different light illumination. Corresponding histogram showing relative enrichment and diameter distribution is plotted in **Figure 2(c)**. Normalized RBM spectra of the SWNTs separated in the supernatant and aggregated floc shows relative enrichment in specific diameter of SWNTs.

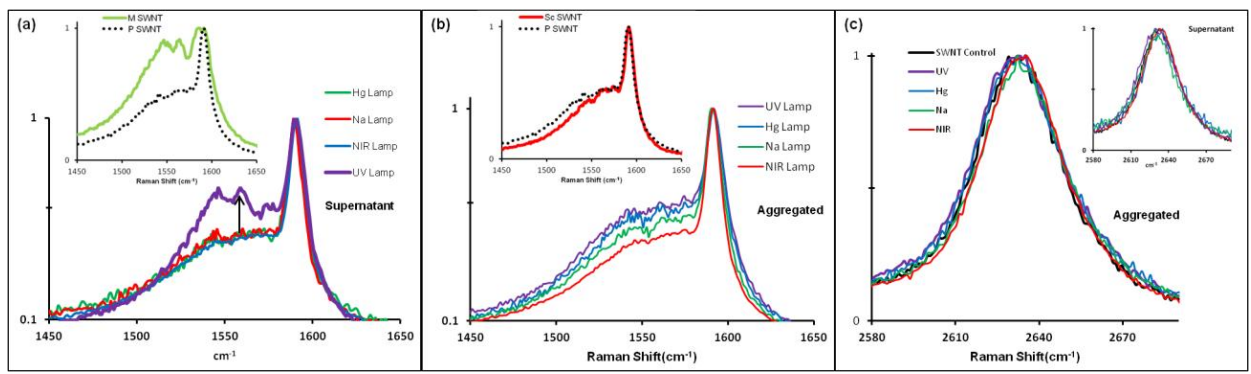

**Figure 3.** G-band Raman spectra of pristine SWNTs under the UV-Hg-Na-NIR lamps. (a) Supernatant. Inset shows controlled pure metallic and pristine SWNT's G-band. (b) Aggregated SWNTs and inset show the G-band of semiconducting and pristine SWNT's. (c) G'-band of aggregated SWNTs (inset supernatant).

G band in SWNTs involves an optical phonon mode between two dissimilar carbon atoms in the unit cell. **Figure 3(a, b)** shows that the G band feature for SWNTs essentially consists of two main components, one peaked at 1592 cm$^{-1}$ (G$^+$) and the other at about 1560 cm$^{-1}$ (G$^-$). The G$^+$ feature is associated with carbon atom vibrations along the nanotube axis (LO phonon mode) whereas, G$^-$ feature is associated with vibrations of carbon atoms along the circumferential direction (TO phonon).[18] The lineshape of G$^-$ band is highly sensitive to the electronic type of the SWNTs. G$^-$ band for metallic SWNTs show up as broadened Breit–Wigner–Fano lineshape whereas Lorentzian lineshape is observed for semiconducting SWNTs.[19] This broadened lineshape of G⁻ feature in metallic SWNTs is related to the presence of free electrons in nanotubes.[20, 21]

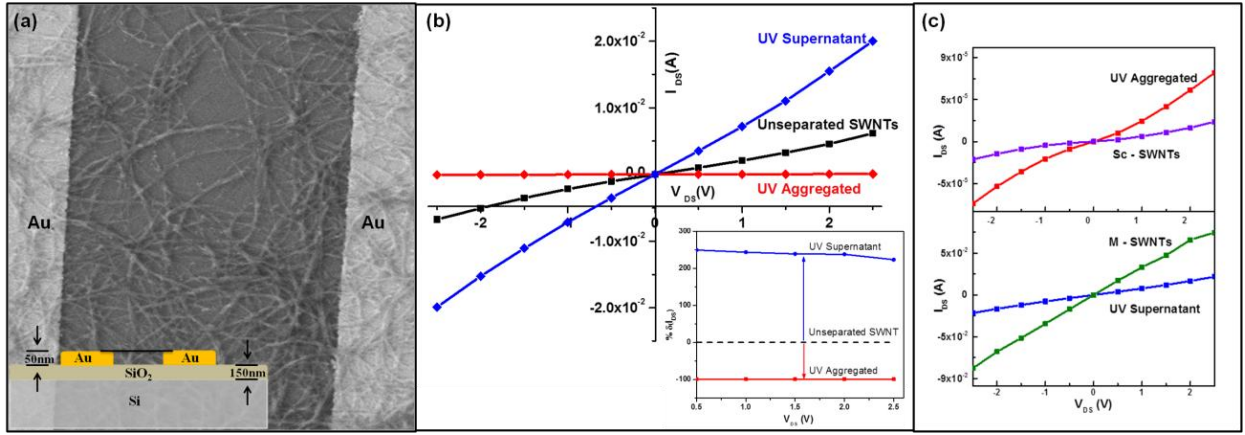

**Figure 4.** (a) SEM image of the fabricated SWNT –FET. [Inset] Representation of the device (b) I-V characteristics of un-separated pristine SWNTs compared with the separated supernatant and SWNTs aggregated by UV light. [Inset] Normalized % Change in current ($\delta I_{DS} = \frac{I_{DS}^{Separated} - I_{DS}^{Unseparated}}{I_{DS}^{Unseparated}} \times 100$) in the optically separated SWNTs as compared with the unseparated pristine SWNTS (c) [Top] I-V characteristics comparing SWNTs aggregated under UV with pre-separated semiconducting SWNTs. [Bottom] SWNTs remaining dispersed in the separated supernatant and pure metallic SWNTs.

SWNTs separated under the UV illumination were further electronically characterized using two probe system (Kiethley 4200-SCS). Field effect transistors based on the supernatant and SWNTs aggregated due to UV illumination and also were based on pre-separated pure (95%) metallic and semiconducting SWNTs were fabricated using direct laser lithography (MICROTECH LW405A) and conventional "lift-off" process. p-doped silicon wafer (purchased from Semiconductor Wafer. Inc) with 150 nm of thermally grown oxide was used as the substrate.

SWNTs separated by the UV exposure were drop casted on the fabricated, 50nm sputter-coated gold patterns. These gold patterns forms the source and drain of the SWNT FET with the sample chuck as the back gate (**Fig 4a**). I-V characteristics of the fabricated SWNT- FET (Fig 4b) based on the SWNTs remaining dispersed in the supernatant of the pristine solution after separation of optically aggregated SWNTs, show distinctly high linear conductivity as compared to unseparated pristine SWNTs. Complimentarily, optically (UV) separated SWNTs in the aggregated floc show considerable decrease in the drain-source current for similar voltages. Inset shows as compared to pristine SWNTs, optically aggregated SWNTs and ones remaining dispersed in the separated solution show considerable decrease (100%) and increase in current (300%) respectively. Moreover, where as SWNTs in the supernatant show linear resistivity indicating enrichment of metallic SWNTs, separated SWNts in the aggregated floc show non linear, diode characteristics as expected for semiconducting SWNTs.

Fabricated FET based on both the SWNTs; ones remaining dispersed in the separated, supernatant solution and ones aggregated out under UV illumination were compared with the similar devices based on reference solution of pre separated, semiconducting and metallic SWNTs (95% Pure Iso-Nanotubes ). SWNTs remaining in the supernatant show similar, linear I-V characteristics as that of metallic Iso-nanotubes (**Figure 4c** [bottom]). Metallic Iso-Nanotubes and supernatant separated under UV illumination show conductivity of 303.456 $Sm^{-1}$ and 73 $Sm^{-1}$ respectively.[22, 23] Whereas, optically aggregated SWNTs show conductivity of 58.7 μS, comparable to 28.7 μS of reference semiconducting SWNTs.( **Figure 4c** [Top]). Hence, the devices based on SWNTs separated by UV illumination show discrete separation of metallic SWNTs in the dispersed supernatant and enrichment of semiconducting SWNTs in the optically aggregated flocculants. Devices based on reference, pre separated semiconducting and metallic

SWNTs also confirms the same. Details of the device fabrication process are provided in S.1. (Supplementary Information)

In conclusion, we report simple, cost effective method of separation of metallic and semiconducting single walled carbon nanotubes of specific diameters using the phenomenon of optically induced aggregation. Moreover, SWNTs separated under lamps of different frequencies show discrete separation and enrichment in specific diameter of SWNTs. Diameter distribution and specific enrichment was analyzed for both the complementary samples, SWNTs remaining dispersed in the solution and ones aggregating out under different lamps. Distinctive enrichment in metallic and semiconducting SWNTs are observed in SWNTs remaining dispersed in supernatant and SWNTs, aggregating under UV illumination. Absorption and Raman spectra, as well as I-V device characteristics all confirm this enrichment of metallic SWNTs in the supernatant and that of semiconducting SWNTs in the separated floc, aggregating out under UV illumination. Results have also been cross verified and compared with pre-separated reference solutions of semiconducting and metallic SWNTs. We report novel phenomenon of optical separation of SWNTs and believe results shown here will be applicable for developing large scale efficient processes for separation of metal/semiconducting as well as different diameter of SWNTs.

ACKNOWLEDGMENTS

Authors are deeply indebted to Ramanujan fellowship (SR/S2/RJN-28/2009) and funding agencies DST (DST/TSG/PT/2012/66), Nanomission (SR/NM/NS-15/2012) for generous grants.